\documentclass{article}
\usepackage{spconf,amsmath,graphicx}

\usepackage{booktabs}
\usepackage{todonotes}
\usepackage{balance}

\title{Federated Learning for ASR based on Wav2vec~2.0}
\name{Tuan Nguyen, Salima Mdhaffar, Natalia Tomashenko, 
Jean-François Bonastre, Yannick Estève\thanks{This work was supported by the French National Research Agency under project DEEP-PRIVACY (ANR-18-CE23-0018), VoicePersonae, H2020 SELMA and was granted access to the HPC resources of IDRIS under the allocation 2021-AD011012551 made by GENCI. }}
\address{LIA - Avignon University, France}
\begin{document}
\ninept
\maketitle
\begin{abstract}
This paper presents a study on the use of federated learning to train an ASR model based on a wav2vec~2.0 model pre-trained by self supervision.
Carried out on the well-known TED-LIUM~3 dataset, our experiments show that such a model can obtain, with no use of a language model, a word error rate of 10.92\% on the official TED-LIUM~3 test set, without sharing any data from the different users.
We also analyse the ASR performance for speakers depending to their participation to the federated learning.
Since federated learning was first introduced for privacy purposes, we also measure its ability to protect speaker identity. 
To do that, we exploit an approach to analyze information contained in exchanged models based on a neural network footprint on an \textit{indicator} dataset.
This analysis is made layer-wise and shows which layers in an exchanged wav2vec~2.0-based model bring the speaker identity information.

\end{abstract}
\begin{keywords}
Federated learning, Automatic Speech Recognition, Self-supervised models, Privacy
\end{keywords}
\section{Introduction}
\label{sec:intro}

Federated learning (FL) has been successfully explored for image and natural language processing \cite{konevcny2016federated}.
FL \cite{konevcny2016federated,mahan16fedavg} is a distributed machine learning paradigm that aims to collaboratively train a machine learning model without data sharing.
It consists in a network of multiple clients and one server.
The training is based on an iterative numbers of rounds.
At each federated learning round, clients train a local model using their private data, and send this updated model to the  server.
The server aggregates the received updates into a single global model and sends its parameters back to the clients' devices.

Recently, FL has been applied in  various speech-related applications, such as automatic speech recognition (ASR) \cite{dimitriadis2020federated,tomashenko2022privacyattacks,gao2022icasspebd2end,zhu22b_interspeech,jia22_interspeech,mehmood22_interspeech}.
keyword spotting \cite{leroy2019federated, hard2020training, hard22_interspeech},
speaker recognition \cite{woubie2021federated,granqvist20_interspeech},
speech emotion recognition \cite{latif2020federated},
% speech emotion recognition \cite{latif2020federated,feng22_interspeech},
self-supervised learning (SSL) of speech representations~\cite{gao22d_interspeech}, and  others.
However, their robustness capabilities have not been extensively investigated and research in this area is still limited. 
In \cite{yu2021federated}, authors  showed why ASR FL task can be considered as very challenging.
The challenges include: (1) communication bottleneck, (2) computation capabilities and energy states, (3) the performance and  accuracy of the learned model and (4) privacy and security considerations.

Recently, wav2vec2.0 \cite{baevski2020wav2vec} models have become more popular and have achieved good performance in many speech processing tasks \cite{yang2021superb,evain2021lebenchmark}.
Authors in \cite{baevski2020wav2vec} claim this framework can enable automatic speech recognition models with just 10 minutes of transcribed speech data.

This paper presents a study on the use of federated learning to train an ASR model based on a wav2vec~2.0 model pre-trained by self supervision.
To our knowledge, there is no previous published work on this use of wav2vec2.0 models.
We analyse the global ASR performance but also the performance for speakers depending to their participation to the federated learning.
Since federated learning was first introduced for privacy purposes, we also measure its ability to protect speaker identity. 
Some related works show that federated learning is vulnerable to various types of attacks \cite{9747231,tomashenko2022privacyattacks}.
To do that, we exploit an approach to analyze information contained in exchanged models based on a neural network footprint on an \textit{indicator} dataset.
This analysis is made layer-wise and shows which layers in an exchanged wav2vec~2.0-based model bring the speaker identity information.

\section{Federated learning for wav2vec~2.0 models}
\label{sec:flasr}

\subsection{Federated learning}

The idea of FL paradigm is about training a neural model across multiple cross-device or server. 
Unlike distributed learning, FL participants only exchange model parameters without exposing any data samples.
By doing this, it is expected to ensure the data privacy of participants or clients. FL technique follows strictly to these steps: \\
\begin{enumerate}
    \item 
The centralized server initializes the global model $G$.
 \item 
 The global model $G$ is sent to each available clients.
 \item 
 Each client $c$ fine-tunes the global model on its local data to obtain the updated model $M_c$.
 \item 
 All the updated models $M_c$ from clients $c_k$ are sent back to the server and being aggregated to form a new model.
 \item 
 The process restarts again from $2^{nd}$ step to $4^{th}$ step until the convergence or number of rounds $T$ is reached.
\end{enumerate}

%\begin{figure}[h!]
%    \centering
%     \includegraphics[scale=0.31]{picture/FL_graph_medium.png}
%     \caption{Federated learning process}
%\end{figure} 

In recent years, more and more studies have been conducted to find the most proper weight aggregation strategy for FL~\cite{Li18Fedprox,Li19qFL}. 
Among them, Federated Averaging (FedAvg)~\cite{mahan16fedavg} is the fundamental and the most well-known FL algorithm. 
FedAvg is based on FedSGD~\cite{sho15fedsgd} algorithm. 
Instead of exchange the gradients after batch updated, FedAvg clients send the updated weights. 
At each round, a number $m$ of clients are chosen among $K$ total clients to send their updated model to server. 
From here, server weights each of clients' parameters $F_k(w)$ by their size of dataset $n_k$ over the total data $n$ are used in the given round and then aggregates them:
% according to the corresponding weight calculated by the ratio of the data it has $n_k$ to the total data $n$ participating in the aggregation process as can be seen in the following equation:
% %\ref{equ:FedAvg}

\begin{equation}
    \begin{split}
        f(w) = \sum_{k=1}^{m}\frac{n_k}{n}F_k(w) 
    \end{split}
\label{equ:FedAvg}
\end{equation}
% % With
% % \newenvironment{conditions}
% %   {\par\vspace{\abovedisplayskip}\noindent\begin{tabular}{>{$}l<{$} @{${}={}$} l}}
% %   {\end{tabular}\par\vspace{\belowdisplayskip}}
  
% % \begin{conditions}
% % $ f(w) $     & global model \\
% % $ n $     &  total dataset's size \\
% % $ n_k $   & dataset's size of client k \\
% % $ F_k(w) $     &  $ \frac{1}{n_k}\sum_i\in \rho_k $ \\
% % $ \rho_k $     &  List of index of data points of client k \\
% % $ f_i(w) $     &  Global model at iteration i
% % \end{conditions}
% \salima{We need I think the formula of aggregation}

\subsection{Wav2vec~2.0 model}

Wav2vec~2.0 \cite{baevski2020wav2vec} is a model pre-trained through self-super\-vision. It takes raw audio as input and computes speech representations. 
Such pre-trained models can be fine-tuned by supervision to build speech recognition systems. 
Wav2vec~2.0 contains three main components: a convolutional feature encoder, a context network and a quantization block.
The convolutional feature encoder converts the raw waveform into a latent representation.
This representation is given to the context network which takes care of the context.
The context network architecture consists of a succession of several transformer encoder blocks.
The quantization network is used to map the latent representation to quantized representation.
Facebook AI released self supervised pre-trained wav2vec 2.0 models.
In this study, we use the model LS960-LV60~\footnote{https://huggingface.co/facebook/wav2vec2-large-960h-lv60} pre-trained on  English.

%\vspace{-0.2m}

\subsection{Implementation with SpeechBrain and Flower toolkits}

The purpose of this research is to train an ASR with a federated learning paradigm.
To attain this objective, \textit{Flower}~\cite{beutel2022flower} and \textit{SpeechBrain}~\cite{ravanelli2021speechbrain} have been used.
Flower is an open-source framework that allows us to build FL experiments and considers the highly varied FL facility scenarios.
The framework is composed of three main components: a client, a server and a strategy.
The client and the server implement the basic functionalities of the clients and a server in a FL training.
%Flower encapsulates a large number of built-in strategies.
%Strategies are algorithms that decide the way to sample clients,  to configure clients for training, to aggregate updates and to evaluate models. 
% 
SpeechBrain is an open source toolkit for automatic speech processing. This toolkit is a simple and flexible end-to-end ASR framework. %SpeechBrain provides many state-of-the-art recipes and pre-trained models for ASR that can be reused.

\vspace{-0.2cm}

\section{Experiments}
\label{sec:exp}
\subsection{Data\label{sec:data}}

The experiments were conducted on the TED-LIUM~3 corpus~\cite{hernandez2018ted} that contains  TED talks with the total  amount of  452 hours of speech data in English from about 2K speakers.
This dataset has been used in some research works in the context of collaborative learning experiments \cite{9747231,tomashenko2022privacyattacks,10.1007/978-3-030-87802-3_39}.
We organized the TED-LIUM~3 training dataset in order to simulate a realistic federated learning framework. 
Each speaker in the TED-LIUM~3 training set acts as a client in this FL scenario. 
For speakers $s$ in the training set with duration $>$ 10 minutes, we consider a  subset of 5 minutes of speech data called $analysis^{s}$ to analyse some FL behaviours.
The remaining is called $train^{s}$ and represents the local dataset for the client.
For speakers in the training set with duration $<$ 10 minutes, all the speaker data will represent the local dataset for the client. 
For the test and development sets, we use the official test and development sets (legacy distribution) of the TED-LIUM~3 release.
The \textit{indicator} dataset~\cite{tomashenko2022privacyattacks} is used to analyse the speaker information contained in the models exchanged  between the server and  clients.
The speakers in the test, development, train, and indicator dataset are disjoint.
Table \ref{tab:dataStat} presents the statistics of the data \footnote{https://github.com/tuanct1997/Federated-Learning-ASR-based-on-wav2vec-2.0}.
%For the reproducibility of experimental results by research community,
%we will make available this partition.}.

\begin{table}[h!]
\vspace{0.3cm}
\label{tab:dataStat}
\centering
\small
\begin{tabular}{l|l|l|l|l|l}
 \toprule
 & \begin{tabular}[c]{@{}l@{}} \textbf{Train}\\ (\textbf{clients})\end{tabular} & \textbf{Analysis} &\textbf{Dev} & \textbf{Test} & \textbf{Indicator} \\  \midrule
Duration, hours &   252.17 & 110.34    &  3.76  &    3.73 & 0.51       \\ 
%1869.66s ~ 31.1m (Indicator), Dev and Test static follow LIUM3 paper , 907842.3700000006 for 1943 clients training set ( Since the training for 1943 clients actually being split to train,dev,test by ratio 80,10,10 so that's why we only have 252.17h of training audio.

$\#$ speakers &   1943  & 1341  &  16    &   16  &    40\\       
$\#$ utterances &    148332 & 65430      &     1155 &   507  & 342   \\ \bottomrule
\end{tabular}
\caption{Data sets statistics}
\end{table}

\subsection{Models \label{sec: model}}

\subsubsection{ASR model based on CRDNN \label{sec:FB}}

In our experiments, we use an attention-based encoder-decoder \cite{chiu2018state} neural network. 
The encoder is a convolutional, recurrent and deep neural network (CRDNN) block.
This CRDNN is composed of three blocks of convolutional neural networks (with respectively 128, 200 and 256 channels) with pooling layers, followed by five bidirectional 1024-dimensional LSTM layers connected to two dense 1024-dimensional layers to  obtain the final acoustic representation. 
The input signal representation is composed of 80-dimensional filter-banks. 
The encoder maps these filter-banks to a higher-level feature representation that are
then passed to a 1024-dimension attention module that identifies which parts of high level speech representations are relevant to each step of the decoding process.
The attention module output is used to compute a context vector used as the input of an RNN decoder. This RNN decoder is composed of 1024-dimensional GRU layer before the output softmax layer. 
The neural network output corresponds to 500 byte pair encoding unigram units, that are a mix between characters and words.
An initial  end-to-end ASR model pre-trained on the CommonVoice dataset\footnote{https://github.com/speechbrain/speechbrain/tree/develop/recipes/\\CommonVoice/ASR/seq2seq} is used to initialize weights for the server model. 
The client's models, trained on the local speaker set for 20 epochs, uses the same model configuration and hyperparameter settings. 

\subsubsection{ASR model based on Wav2vec~2.0 \label{sec:w2v}}

The architecture is based on an end-to-end approach with SSL.
The system  is composed of the large pre-trained English wav2vec~2.0 model, a linear layer of 1024 units, and the softmax output layer.
 Connectionist temporal classification (CTC) loss function \cite{graves2006connectionist} is used as a loss function.

We conduct our experiments using randomly initialized weights for the server model. 
The final model is trained on 5 V100 GPUs, each with 32 GB memory, for 100 rounds at a batch size of 4. 
The clients' models, trained on the local speaker set for 20 epochs, use the same model configuration and hyperparameter settings.

\subsection{FL for ASR}
With the data set split described in Section~\ref{sec:data}, FL's clients mostly cover only a small amount of audio. This is a big challenge to train an ASR model since normally, in this domain the dataset size is a critical point. Moreover, between these clients, a big difference in voice, audio quality, utterances or data size leads to an extreme non independent and identically distributed (non-IID) case. As being seen in different studies~\cite{zhao18iid}\cite{huang22silo}\cite{gao2022icasspebd2end}, non-IID is a big challenge for FL. 
For end-to-end ASR, Yan Gao et al \cite{gao2022icasspebd2end} also found that it is nearly impossible to start the training from scratch. 
In their work, they started the training from a pre-trained model on half of same dataset. Described in Section \ref{sec: model}, for both CRDNN and wav2vec 2.0 architectures, a pre-trained model is used as an initialised global model. The difference is that for CRDNN, the pre-trained model is already specialised for the ASR task. 
For wav2vec 2.0, the initialised model is only pre-trained to learn the representation from audio of LibriSpeech data.

To set up a FL ASR experiment, a simulation has been created and executed on the same machine. Within the simulation, we have 1 server and 1943 clients corresponding to speakers in the TED-LIUM 3. In a normal FL scenario, as can be seen in equation~\ref{equ:FedAvg}, only $m$ out of $K$ clients will participate in aggregation process because $K$ can be very large and using all $K$ clients can be unmanageable.  
In our experiments, 20 clients per round are chose to participate for both CRDNN and wav2vec 2.0 architectures. 
Indeed, during our experiments we found that $m=20$ is the best trade-off point for our resource\footnote{To facilitate research in FL using TED-LIUM~3 dataset, the recipe including data preparation, FL training and evaluation scripts will be released open source upon paper acceptance}.

\subsection{ASR performance}

\subsubsection{General performance \label{sec:general_res}}
\begin{figure}[h!]   
    \centering
    \includegraphics[width=\linewidth]{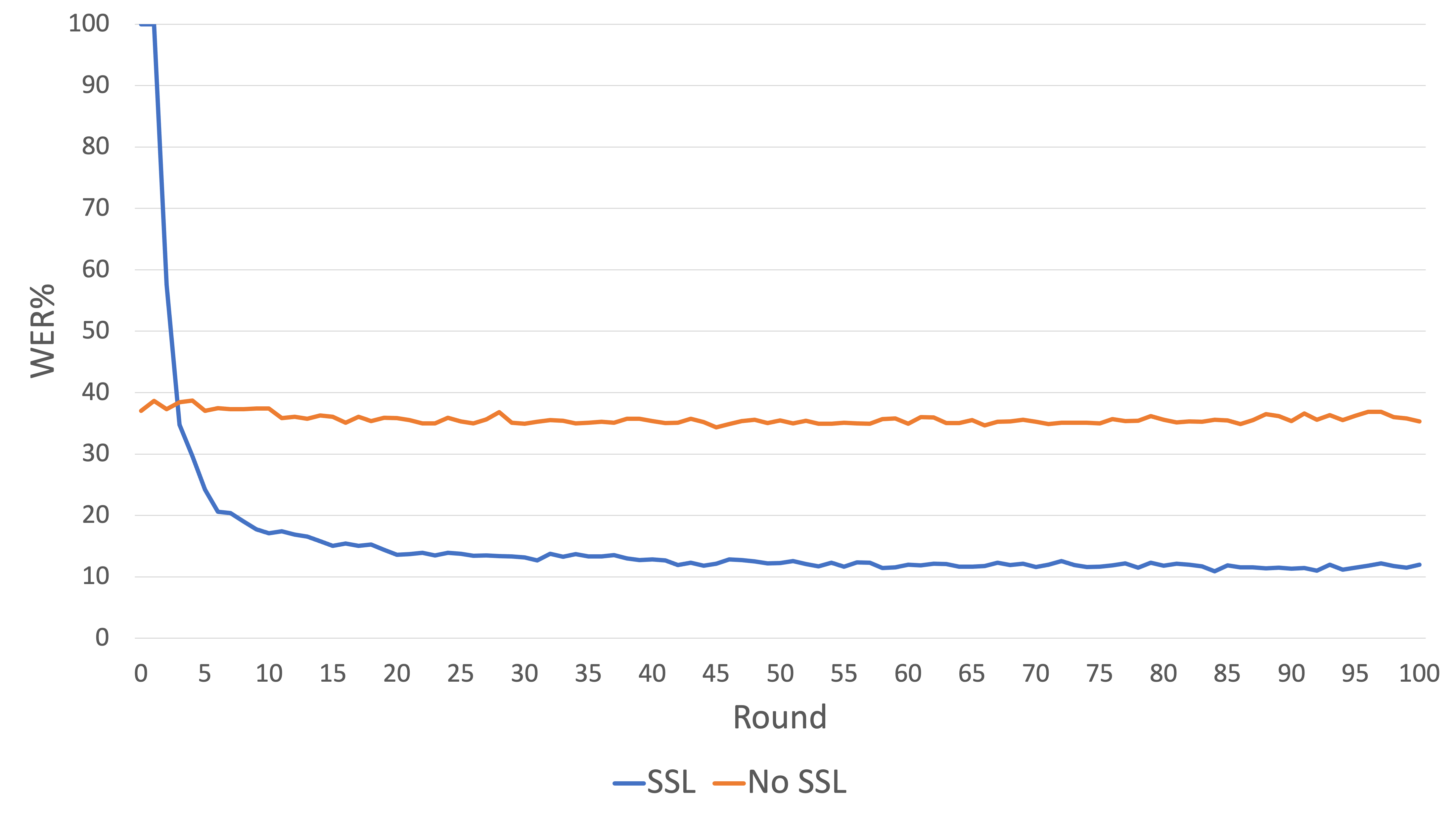}
    \caption{Performance evolution (WER,\%) of end-to-end models ASR based on wav2vec~2.0 (SSL) compared to  end-to-end ASR models  based on CRDNN (no SSL) on the  TED-LIUM~3 test dataset
    }
    \label{fig:wer_FB_W2V}
\end{figure}

To analyze the performance of the FL ASR system, the global model was tested on the TED-LIUM 3 test set where speakers of this set were never exposed during the training phase.
Results in Figure \ref{fig:wer_FB_W2V} show that it is possible to improve the ASR performance  in terms of word error rate (WER) for  speakers unseen during the FL training (speakers in the test set).

Having a better start in terms of WER (WER = 37.04\% at first round), CRDNN still struggles to converged compared to wav2vec 2.0. 
%WER 
% Despite CommonVoice pre-trained model is good at ASR task, we noticed this is not enough since there is a too large gap between CommonVoice and TED-LIUM~3. 
Despite the CommonVoice pre-trained model is good at the ASR task, we noticed this is not enough since there is a  large gap between CommonVoice and TED-LIUM~3. 
WER of CRDNN stays at around $35\%$ for the rest of rounds (best performance is reached at round 45 WER = 34.33\%).

By using only a small dataset at local level, it's not enough to learn the information of TED-LIUM 3 using the CRDNN pre-trained models. 
On the other hand, with wav2vec 2.0 the problem seems to be overcome. 
Just within 4 rounds, wav2vec 2.0 caches up with CRDNN and keeps improving. 
The best performance is recorded at  $85^{th}$ round with $10.92\%$ of WER. 
In addition, for 100 rounds, the FL is only contributed by 1209 speakers (which equal to $62\%$ of total speakers) to reach this performance.

\subsubsection{Longitudinal speaker-level ASR performance \label{sec:asrperformance}}

Our hypothesis is that performance of FL in each round is still affected by the speaker's participation in each round.
At each round, a fixed number of clients participate in the training by sending to the server a model trained using their private data, this may  
cause  \textit{forgetting} of previously-learnt knowledge related to the  speakers seen in previous rounds.
In contrast, in a centralized training process, all speaker data is used simultaneously, eliminating the risk of forgetting previously learned speaker information.

To address this aspect, we propose to analyze the evolution of WER per speaker according to the different rounds.
As described in Section \ref{sec:data}, 5 minutes of speech have been removed from the training data for some speakers and included into an $analysis$ dataset.

Let us denote $G_{r}$ the general model at communication round $r$ and consider the speakers that share their models during the FL run at round 5.
To facilitate the analysis, we pick a subset of 5 speakers at this round to test the performance of $G_{r}$.

\begin{figure}[h!]
    \centering
    \includegraphics[width=\linewidth, height=5cm]{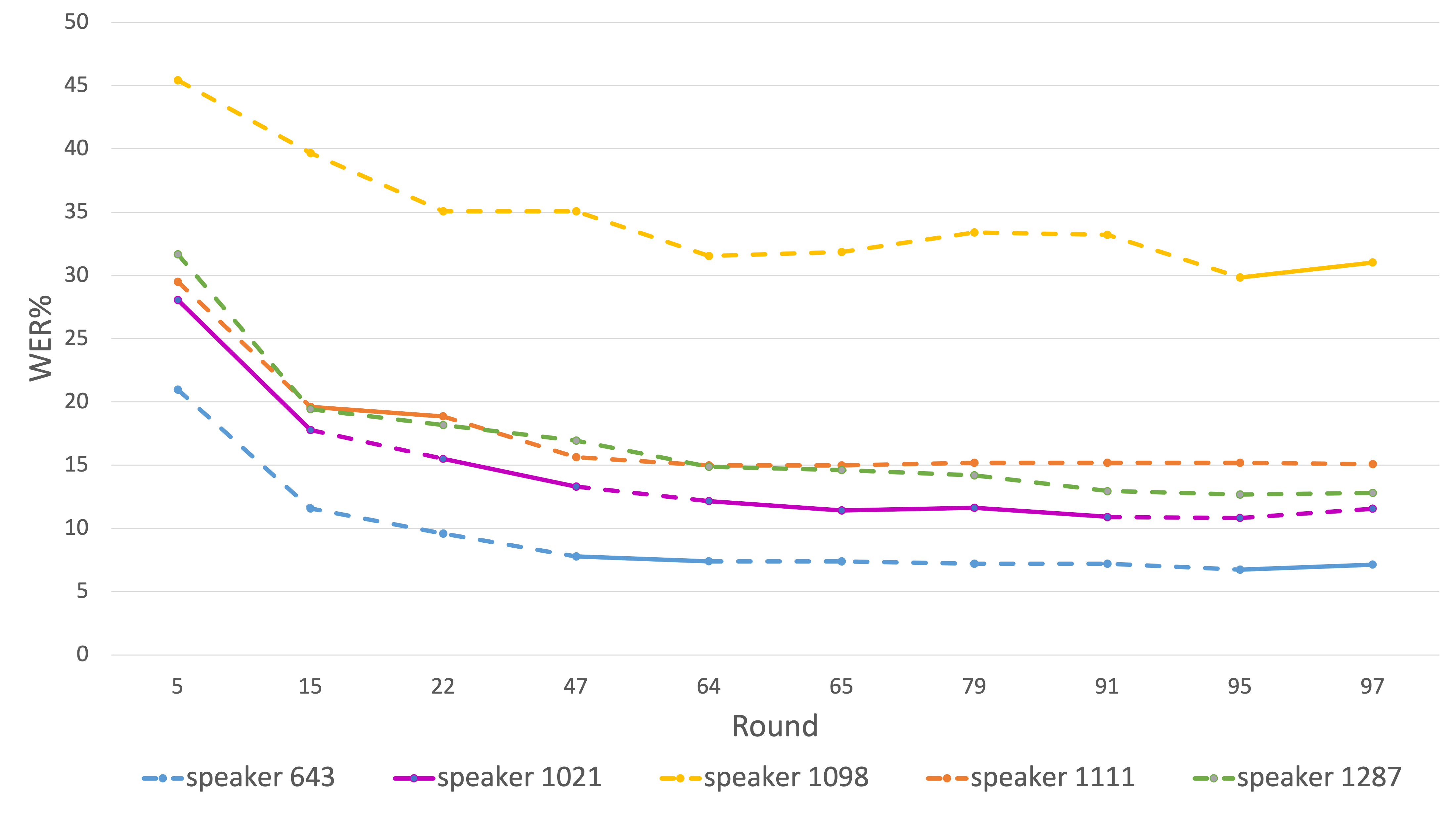}
    \caption{Global model performance on speaker's \textit{analysis} dataset }
    \label{fig:wer_seen_speaker}
\end{figure}

\begin{figure}[h!]
    \centering
    \includegraphics[width=\linewidth, height=5cm]{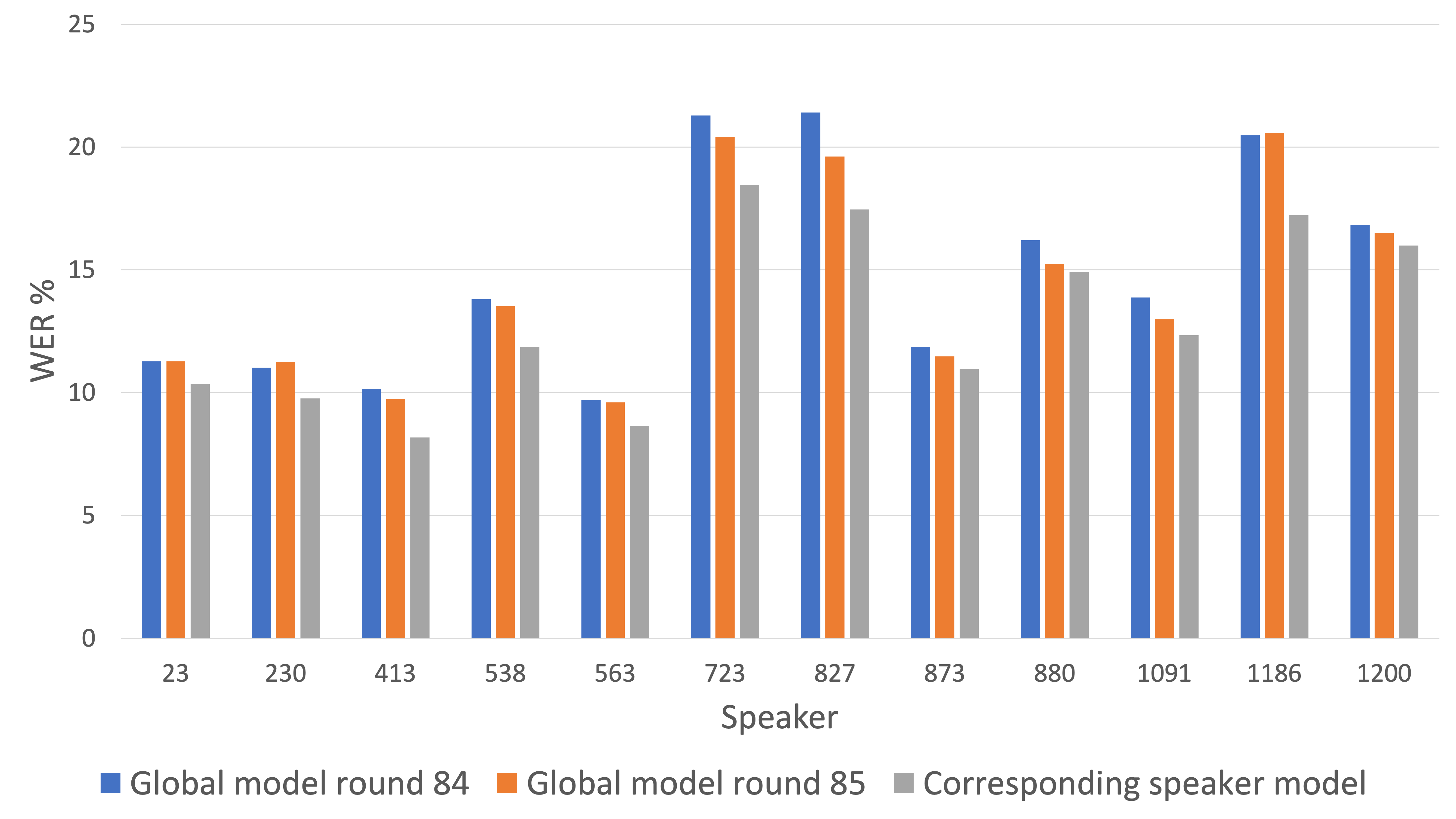}
    \caption{Performance of speaker local models on corresponding \textit{analysis} datasets compared to the global model}
    \vspace{-0.5cm}
    \label{fig:speaker_performance}
\end{figure}

Figure~\ref{fig:wer_seen_speaker} shows how the general model $G_{r}$ performs on these five \textit{analysis} datasets at each round.
A dotted segment means that the speaker was not involved between two rounds, while a solid line means that the speaker contributed for one round during the two ones connected by the segment.
First, we observe similar trends between different speakers and between solid and dotted segments.
Figure~\ref{fig:wer_seen_speaker} shows a significant improvement in  WER for first rounds (from round 5 to round 22). 
Then, for next rounds, WER  for different speakers does  not vary significantly.
Therefore it can be seen that $G_{r}$ contains relevant information for these speakers and does not bring any bias based on the number of participations.

Another important aspect of FL to be considered is the performance of a speaker's local model. Figure~\ref{fig:speaker_performance} reports a test performed using the best round (round $85^{th}$). 
The figure shows that the local speaker model (represented by the gray columns) is enhanced after fine-tuning on its own dataset, as expected.
Then we tested all these speaker models on the TED-LIUM~3 test set and obtained the average WER $= 13.04\%$, to be compared with WER $= 10.92\%$ of global model as been reported in Section \ref{sec:asrperformance}.
These results show that the global model $G_{r}$ is well designed to process not only speakers involved in the federated learning but also  new speakers.

\subsection{Protection of speaker identity}

\subsubsection{Privacy preservation scenario and attack model}\label{sec:attack}

Privacy preservation can be formulated as a game between \textit{users} who share some data and \textit{attackers} who access this data or data extracted from it and want to get information about the users~\cite{Tomashenko2021CSl,tomashenko2020introducing}.
In FL,   to  preserve the user's data, only model updates are transmitted between the clients and server. An attacker aims to attack users using information received  from the server about  the models' updates.

In this work, we consider the following privacy preservation scenario.
We assume that  an attacker  has access to the following data and models: (i) a global model $G_r$ shared with the clients at communication round $r$; (ii) a personalised model $M$ of some speaker obtained at round $r$ from  $G_r$; (iii)  speech  data (utterances $u_1,\ldots,u_T$) of a known speaker  which will be referred to as  \textit{enrollment} data following the traditional ASV terminology data.
% \footnote{\label{fn:1}following the traditional ASV terminology} data.
The attacker does not know the identity of the speaker corresponding to $M$ and aims to perform an automatic speaker verification  (ASV) task using i--iii  to verify if the model   $M$ and data $u_1,\ldots,u_T$ correspond to the same speaker. We will refer to  $M$ as \textit{test trial} model. 
%\footref{fn:1} 

In this work, we use  an attack model that is similar to the one proposed in \cite{tomashenko2022privacyattacks}.
The idea is based on capturing  information about the speaker identity from the corresponding personalised model $M$ and the global model $G_r$  
by comparing the outputs of these two neural acoustic models (AM) taken from  hidden layers $h$ on some speech data (called  \textit{indicator} in \cite{tomashenko2022privacyattacks}). The  \textit{indicator}  data is not related to test or  training data and can be chosen arbitrarily from any speakers. 

The method consists in the following steps: 
(1) get a personalised model $M_e$  for enrollment speaker $e$   from the enrollment data (iii): $u_1,\ldots,u_T$, by finetuning  $G_r$ on this data;
(2) using $G_r$ and $M$ compute per-frame differences between  activation values of these two models from some hidden layer $h$ for all utterances of the indicator dataset; then  for these differences compute a mean vector $\mu$ over all frames; 
(3)   using $G_r$ and $M_e$ compute per-frame differences between  activation values of these two models from layer $h$ for all utterances of the indicator dataset; then  for these differences compute a mean vector $\mu_e$ over all frames; 
(4) compute similarity score $\rho$ between enrollment model $M_e$ and test trial model $M$  as  cosine similarity between corresponding mean vectors: 
$\rho(M_e,M)=\cos(\mu,\mu_e)$ and perform an ASV task using these scores.
More details can be found in~\cite{tomashenko2022privacyattacks}\footnote{ The differences with respect to the work~\cite{tomashenko2022privacyattacks} are in the way the similarity scores are computed: (1) only mean values (without standard deviation components  as in~\cite{tomashenko2022privacyattacks}) are used; and (2)  cosine distance instead of Euclidean-based metric is applied.}.

As a privacy evaluation metric, in this work, we use  equal error rate (EER).
Denoting by $P_\text{fa}(\theta)$ and $P_\text{miss}(\theta)$ the false alarm and miss rates at threshold~$\theta$, the EER corresponds to the threshold $\theta_\text{EER}$ at which the two detection error rates are equal, i.e., $\text{EER}=P_\text{fa}(\theta_\text{EER})=P_\text{miss}(\theta_\text{EER})$.
The higher EER the better is privacy preservation.

\begin{figure}[h!]
    \centering
    \includegraphics[width=\linewidth]{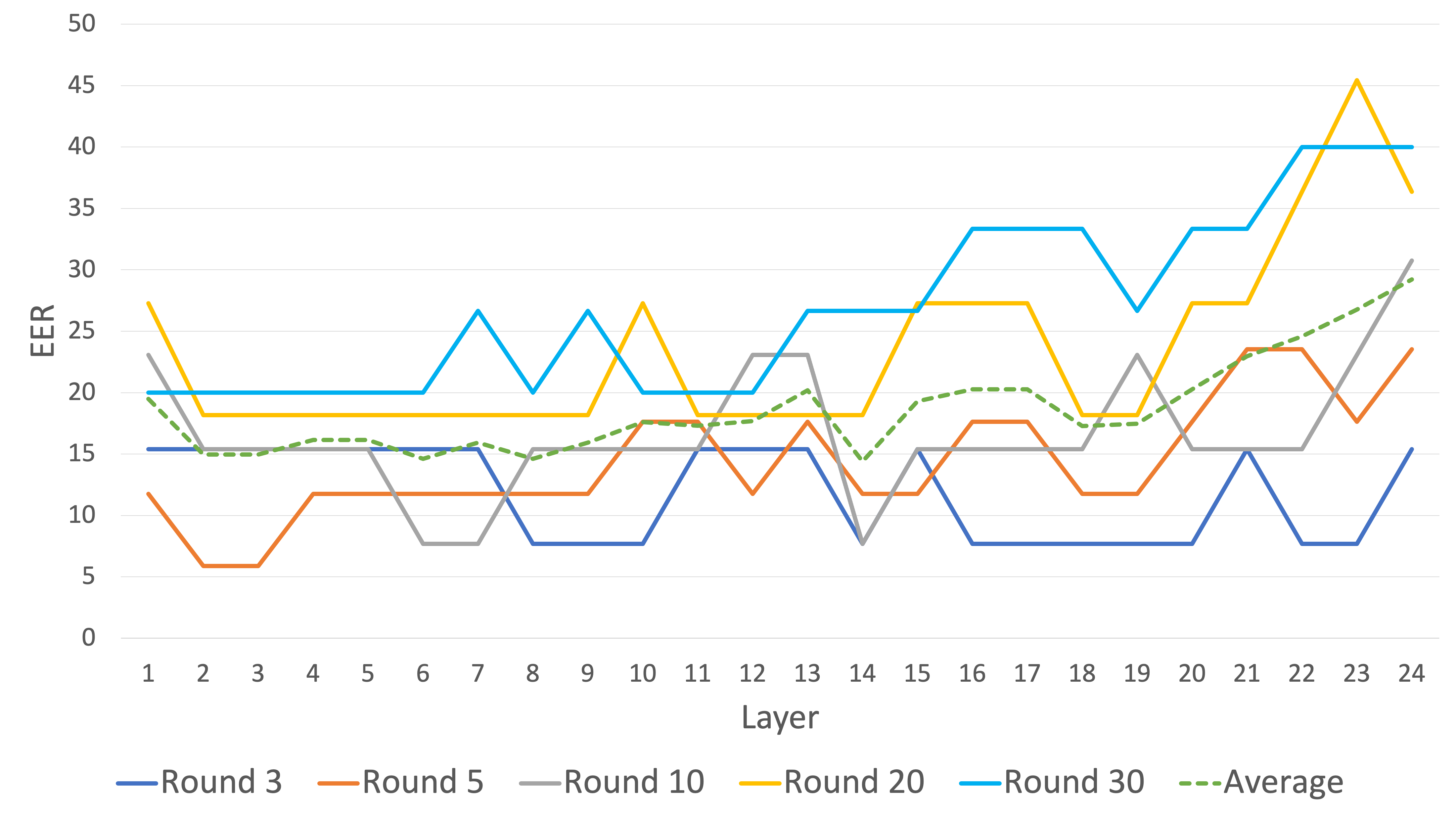}
    \caption{Privacy evaluation (EER,\%) for different computational rounds and layers of the ASR models}
    \label{fig:eer}
\end{figure}

\subsubsection{Results}

The speaker privacy  has been evaluated for the  ASR  models  with the wav2vec~2.0-based architecture (Section~\ref{sec:w2v}). 
We applied the attack model described in Section~\ref{sec:attack} for  different computational rounds~$r$. 
For each round, we used 50 enrollment speakers  and perform an ASV task for all clients of the given round (performing in average  15 target  and 986 non-target trials per round). The amount of enrollment data for each model ((iii) in Section~\ref{sec:attack}) is about 5 minutes.
Experimental results are presented in Figure~\ref{fig:eer} for different hidden layers of the ASR models and  rounds \{3,5,10,20,30\}. 
The green dashed curve represents the EER averaged over selected rounds.
% A STATISTICAL SIGNIFICANCE TEST FOR PERSON AUTHENTICATION: https://publications.idiap.ch/attachments/reports/2004/bengio_2004_odyssey_stats.pdf
% The 95\%-confidence intervals for this curve is [4.8
% 4.3
% 4.3
% 2.5
% 2.5
% 3.9
% 5.6
% 3.9
% 5.6
% 5.6
% 1.5
% 3.4
% 3.6
% 6.3
% 5.6
% 7.9
% 7.9
% 7.7
% 6.1
% 7.9
% 6.1
% 10.7
% 12.3
% 7.8]
In general, EER increases when the number of computational rounds increases, so it is  more difficult for the attacker to retrieve information about the speaker identity from the personalised models on the later rounds than on the earlier ones. 
For lower hidden layers, the EER in average is lower than for upper layers.

\section{Conclusion}

This paper presents a study on the use of federated learning to train an ASR model based on the wav2vec 2.0 model.
The experimental results, carried out on the well-known TED-LIUM~3 dataset, showed the capability of federated learning to train an effective ASR model without sharing any speech data when federated learning is applied to fine-tune a wav2vec~2.0 model.
Our experiments demonstrated that the general model contains relevant information for those speakers who have participated in the federated learning by sharing their local models, but we did not observe any bias based on the number of participations.
The general model built through federated learning is also very effective to process unseen speakers.
Finally, we have evaluated the  privacy level achieved for the proposed federated learning framework by exploiting an approach to analyse information contained in personalised models based on a neural network footprint on an indicator dataset. The layer-wise analysis has  demonstrated that speaker information can be retrieved from all the considered rounds of the FL process. EER is lower on the earlier stages of the process  and varies from 5 up to 20\%, for different rounds for hidden layers  \#2--\#6).
In a future work, we could also investigate which amount of linguistic information is brought by the shared  local speaker models.

%\newpage
% \ninept
%
\balance
\bibliographystyle{IEEEbib}
\bibliography{strings,refs}

\end{document}